\documentstyle[11pt,moriond,epsfig]{article}

\def\Journal#1#2#3#4{{#1} {\bf #2}, #3 (#4)}

\def\be{\begin{equation}}
\def\ee{\end{equation}}
\def\bea{\begin{eqnarray}}
\def\eea{\end{eqnarray}}

\newcommand{\mic}{\,$\mu$m }
\newcommand{\micpa}{\,$\mu$m}

\begin{document}
\def\gtapp
{\mathrel{\hbox{\raise0.3ex\hbox{$>$}\kern-0.8em\lower0.8ex\hbox{$\sim$}}}}
\def\ltapp
{\mathrel{\hbox{\raise0.3ex\hbox{$<$}\kern-0.75em\lower0.8ex\hbox{$\sim$}}}}
\vspace*{4cm}
\title{A ``UV+IR'' HISTORY OF STAR FORMATION AT 0\,$\ltapp$\,$Z$\,$\ltapp$\,1}

\author{ E. LE FLOC'H \& the MIPS team}

\address{Steward Observatory, University of Arizona, 933 N. Cherry Avenue, \\
Tucson, AZ 85721, United States}

\maketitle\abstracts{The combination of both contributions from the
  observed UV emission and the absorbed radiations reprocessed in the
  infrared represents the ideal approach to constrain the activity of
  massive star formation in galaxies.  Using recent results from GALEX
  and {\it Spitzer}, we compare the evolutions of the UV and IR energy
  densities with redshift as well as their contributions to the star
  formation history at 0\,$\ltapp$\,$z$\,$\ltapp$\,1. We find that the
  comoving IR luminosity is characterized by a much faster evolution
  than seen in the UV. Our results also indicate that $\sim$\,70\% of
  the star-forming activity at $z$\,$\sim$\,1 is produced by the
  so-called IR-luminous sources (L$_{\rm
    IR}$\,$\geq$\,10$^{11}$\,L$_\odot$).  }

\section{Dust extinction and the limitations of the UV window}

The multi-wavelength deep surveys performed in the last decade as well
as the detection of the Cosmic Infrared (IR) Background by COBE
revealed the dramatic effects of dust extinction in the distant
Universe. This significant high-redshift reprocessing of
short-wavelength radiations (e.g., UV, optical) into the thermal
infrared appears to be associated very closely to the strong evolution
that IR-luminous galaxies (i.e., L$_{\rm
  IR}$\,$\geq$\,10$^{11}$\,L$_\odot$) have undergone with lookback
time.  The luminosity of such objects is mostly powered by
highly-embedded star formation or by the accretion of dusty material
around active nuclei. As a result they emit the bulk of their energy
between 8 and 1000\micpa.  It is now well established that they
contributed significantly to the assembly of the present-day stellar
mass and to the growth of supermassive black holes (e.g., Blain et
al. 1999, Chary \& Elbaz 2001).

This dominant contribution of IR-luminous systems in the building of
structures simply demonstrates not only that the amount of radiations
absorbed by dust can not be neglected when quantifying the cosmic
evolution, {\bf but also that this dust extinction can not be properly
  quantified from the slope of the UV continuum} as it has been
proposed in the past few years. Based on a local sample of luminous
and ultra-luminous infrared galaxies observed with the STIS camera
on-board the {\it HST}, Goldader et al. (2002) have shown that the
characteristic IR excess of these sources (defined as the ratio
between their IR and UV luminosities) is significantly larger than
what can be predicted based on the attenuation of their UV continuum
and the relation initially proposed by Meurer et al. (1999). This can
be explained by the existence of optically-thick dusty environments
that contribute to a non negligible fraction of the total energy
output of these luminous galaxies but where the UV radiations are
almost completely hidden and absorbed by dust.  It can also be due to
the absence of physical correlation between the dusty regions emitting
the bulk of the bolometric luminosity and the other unobscured
components dominating the UV emission. This has been observed in local
IR-luminous systems resulting from the interaction of a UV bright
source with another red and dusty galaxy (Charmandaris et
al. 2004). As an illustration, we show in Fig.\,1 the cases of VV\,114
and Arp\,299, which clearly reveal a mis-connection between the
regions where are concentrated the IR and the UV emission. Such
systems, if located at $z$\,$\gtapp$\,1, would barely be resolved. The
measured UV continuum would obviously originate from the only
contribution of the UV-bright and unabsorbed components, leading to a
distorted picture of the IR properties of these objects.

\begin{figure}
\begin{center}
\psfig{figure=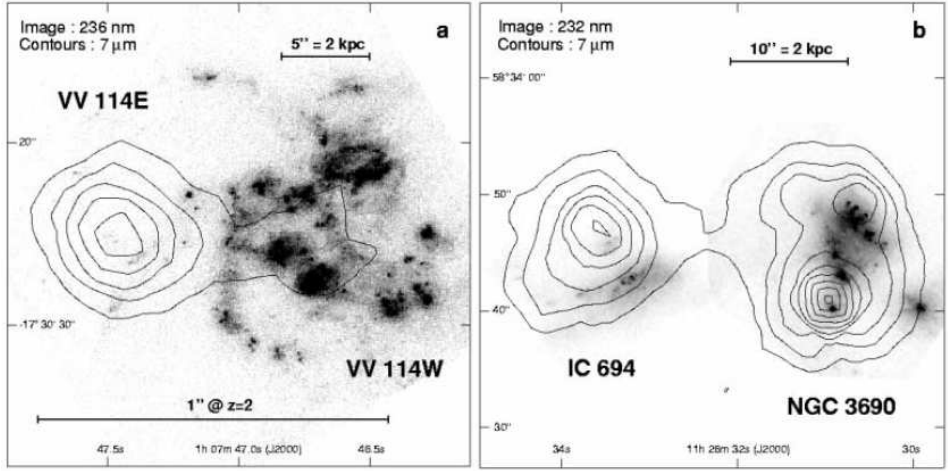,height=2.5in}
\end{center}
\caption{a) An HST/STIS image of VV\,114 at 0.23\,$\mu$m, adapted
  from Goldader et al. (2002) with an overlay of the 7$\mu$m emission from
  ISOCAM. The UV flux limit is
  26.6\,mag\,arcsec$^{-2}$. The contour levels are set with
  logarithmic spacing between 0.8 and 8.4\,mJy\,arcsec$^{-2}$ (
  1$\sigma$ $\sim$ 0.15\,mJy\,arcsec$^{-2}$). b) An archival HST/FOC
  (0.22\,$\mu$m) UV image of Arp\,299 with the 7$\mu$m emission
  contours. The UV flux limit is 21\,mag\,arcsec$^{-2}$. The contour limits are set
  to 1.2 and 30.7\,mJy\,arcsec$^{-2}$ (1$\sigma$ $\sim$
  0.27\,mJy\,arcsec$^{-2}$).
  One can easily see
  that in both systems the rest-frame UV light from the mid-IR
  dominant source is either completely suppressed (VV\,114) or
  marginally detected (Arp\,299). Adapted from Charmandaris, Le Floc'h \& Mirabel (2004).
\label{fig:radish}}
\end{figure}

Recent observations
from the {\it Spitzer Space Telescope\,} actually  show a similar result regarding IR-luminous 
objects at high redshift (Papovich et al., in prep.). This clearly reveals the
severe underestimate of dust extinction (and therefore bolometric luminosity) that can affect
studies of high redshift galaxies when relying on the UV and optical wavelengths only.

\section{Estimating the star-forming activity  of galaxies from the contributions of their
 UV and IR luminosities}

The UV photons produced by massive stars and redshifted into the
optical window allow us to trace on-going starburst episodes within
distant galaxies.  They are nevertheless very efficiently absorbed by
dust grains, and we showed in the previous section that a correct
estimate of this dust extinction can not be derived solely from the
optical wavelengths. Consequently, the cleanest approach to infer the
star-forming activity of galaxies is to account for both the energy
observed in the rest-frame UV and the dust-reprocessed light emerging
in the far-infrared (e.g., Adelberger \& Steidel 2000, Bell 2003). We
present hereafter an estimate of the star formation history at
0\,$\ltapp$\,$z$\,$\ltapp$\,1 based on this approach.

\subsection{Contribution from the observed UV luminosity}

The GALEX satellite now provides new opportunities for probing the
high redshift Universe with great sensitivity in the far-UV. Based on
surveys conducted with this new facility, Arnouts et al. (2005) and
Schiminovich et al. (2005) have derived the evolution of the
luminosity function and the evolution of the luminosity density at
rest-frame 1\,500\,\AA.  They detect a clear increase which follows an
evolution in (1+$z$)$^{2.5\pm0.7}$ at
0\,$\ltapp$\,$z$\,$\ltapp$\,1. Assuming standard conversions from UV
to star formation rate (SFR, Kennicutt 1998), they have translated
their results into an estimate of the contribution of the observed UV
luminosity to the star formation history. It is consistent with the
initial measurements derived by Madau et al. (1996) and Lilly et
al. (1996), and it is represented by the dashed line in Fig.\,2.a.

\subsection{Contribution from the dust-reprocessed thermal IR emission}

In addition to GALEX, the successful launch of the {\it Spitzer Space
  Telescope\,} in August~2003 opened new perspectives for our
understanding of IR galaxy evolution.  {\it Spitzer\,} operates
between 3.6 and 160\mic with unprecedented sensitivity and better
spatial resolution compared to previous infrared satellites (e.g.,
{\it IRAS,\,} {\it ISO}).

In the context of our Guarantee Time Observer program, we have imaged
several cosmological fields ({\it ``Chandra Deep Field South'' (CDFS),
  ``Lockman Hole'', ``Hubble Deep Field North'', ``NOAO Deep
  Wide-Field Survey''}, ...) with the MIPS instrument (Rieke et
al. 2004). MIPS is the {\it ``Multi-Band Imaging and Photometer''\,}
on-board {\it Spitzer}. This instrument can cover large sky areas with
high efficiency, simultaneously acquiring data at 24, 70 and
160\micpa. We have cross-identified our 24\mic observations of the
CDFS with various libraries of spectroscopic and photometric redshifts
published in the literature. Within an area of
$\sim$\,775\,arcmin$^2$, this allowed us to build a sample of
$\sim$\,2\,600 24\micpa-selected galaxies identified with a reliable
redshift at 0\,$\ltapp$\,$z$\,$\ltapp$\,1 (see Le Floc'h et al. 2005).
We found that this sample is mostly complete up to
$z$\,$\sim$\,0.8. Assuming that the usual IR spectral energy
distributions characterizing the IR properties of local galaxies are
still valid in the distant Universe (Elbaz et al. 2005), we derived
the following:

\begin{itemize}
\item{Total IR luminosities}
\vspace{-.1in}
\item{Correlation between optical and IR luminosities}
\vspace{-.1in}
\item{IR to UV ratio as a function of IR luminosity}
\vspace{-.1in}
\item{Stellar masses as a function of IR luminosity}
\vspace{-.1in}
\item{IR luminosity function}
\vspace{-.1in}
\item{Evolution of the comoving IR energy density and equivalent SFR}
\end{itemize}

In particular, we showed that the IR luminosity density has undergone
a strong evolution in (1+$z$)$^{3.9\pm0.4}$ at
0\,$\ltapp$\,$z$\,$\ltapp$\,1. Assuming the standard calibration
between star formation rate and IR luminosity (Kennicutt 1998), we
finally derived an equivalent dusty star formation history illustrated
by the solid line in Fig.\,2.a.

\subsection{The ``UV+IR'' history of star formation}

\begin{figure}
\begin{center}
\psfig{figure=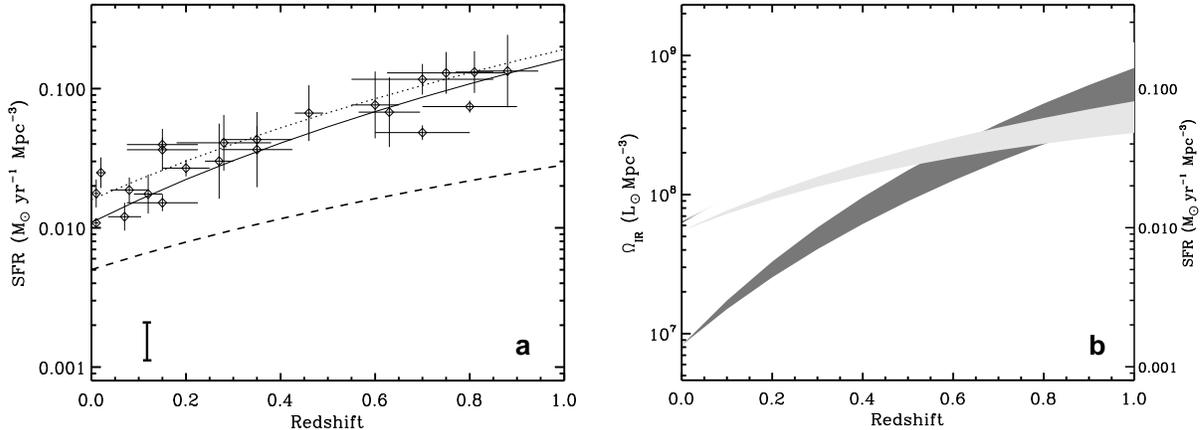,height=2.5in}
\end{center}
\caption{a) Evolution of the observed UV and IR luminosities
 at 0\,$\ltapp$\,$z$\,$\ltapp$\,1 
(respectively dashed and solid lines)
converted to densities of
star-forming activity using the calibrations from Kennicutt (1998). 
 The vertical bar in the bottom left corner
represents their typical uncertainty at $z$\,$\sim$\,1. The 
total star formation history is estimated as the addition of these two contributions
and it  is represented by the dotted line. Data points from 
the literature are also indicated for comparison (open diamonds, see Hopkins 2004 for references).
b) Comparison between the evolutions of the comoving IR energy density produced
by IR-luminous sources (i.e., L$_{\rm IR}$\,$\geq$\,10$^{11}$\,L$_\odot$, 
dark shaded region) and  galaxies with  L$_{\rm IR}$\,$\leq$\,10$^{11}$\,L$_\odot$ 
(light shaded region).
   Adapted from  Le Floc'h et al.  (2005).
\label{fig:radish}}
\end{figure}

As pointed out previously, the proper estimate of the star-forming
activity in a given redshift bin must be obtained by taking into
account both the radiations absorbed by dust and the UV~photons 
directly escaping the HII regions. The corresponding addition
of the observed-UV and dusty-IR star formation histories is
represented by the dotted line in Fig.\,2.a. For comparison, we also
show integrated star formation rate densities and their uncertainties
estimated within various redshift bins and taken from the literature
(see the compilation by Hopkins 2004 for references).

\subsection{Some caveats}

In the conversion between the comoving IR luminosity and the 
dusty star formation rate density, we have assumed that 100\% of the IR energy
was powered by star formation. The contribution of active
galactic nuclei (AGNs) within our mid-IR selected sample has been
therefore neglected. We believe that such contamination should not be
larger than 15-20\% up to $z$\,$\sim$\,1. This estimate is roughly based on
the number of objects that we identified as AGNs using the multi-wavelength surveys
of the CDFS found in the literature. It is consistent with  estimates derived
by other groups (Fadda et al. 2002, Franceschini et al. 2005).

It should also be noted that the uncertainties in the evolution of the
UV and the IR comoving energy densities are not negligible. This
originates from the unconstrained faint-end slope of the luminosity
function at high redshift, and the subsequent degeneracy that appears
when quantifying its evolution. In spite of these uncertainties
however, the difference between the IR and the UV densities
is sufficiently large for being statistically significant.

\section{Discussion}

Figure\,2.a reveals that the {\it relative\,} contribution of the uncorrected
UV light to the total star formation is decreasing
with redshift. This is a direct consequence of the stronger evolution
of the IR energy density (in (1+$z$)$^{3.9\pm0.4}$) 
compared to the evolution of the UV
luminosity (in (1+$z$)$^{2.5\pm0.7}$). 

This strong evolution observed at infrared wavelengths is not due to a
global increase of the density of galaxies across the full range of
luminosities. The IR luminosity functions (LF) that we derived from
the CDFS (Le Floc'h et al. 2005) rather reveal a significant evolution
of the characteristic luminosity L$^*_{\rm IR}$ describing the knee of
the LF. This is consistent with the analysis that have been infered
from the ISO and SCUBA surveys, and it shows that the evolution has
been in fact driven by a strong increase of the relative contribution
from the most IR-luminous sources to the total IR energy output.  We
estimate that such objects were dominating the star formation history
at $z$\,$\gtapp$\,0.7, representing $\sim$\,70\% of the star formation
rate density at $z$\,$\sim$\,1. This is illustrated on Fig.\,2.b,
which compares the evolution of the comoving IR luminosity densities
for IR-luminous sources (i.e., L$_{\rm
  IR}$\,$\geq$\,10$^{11}$\,L$_\odot$) and fainter galaxies with
L$_{\rm IR}$\,$\leq$\,10$^{11}$\,L$_\odot$.

From the GALEX observations, Schiminovich et al. (2005) have noted
that the population of the most UV-luminous galaxies (UVLGs) has also
experienced a dramatic evolution with lookback time. Investigating the
relationship between these UVLGs and the IR-luminous sources should
give us more clues to better understand the origin of the characteristic
decline of the star formation history from $z$\,$\sim$\,1 down to $z$\,=\,0.

We finally note that, according to our current knowledge, the IR
emission alone provides a rather good estimate of the total star
formation history even at low redshifts.  Indeed, the uncertainties
affecting the current estimates (from the IR or the other multi-wavelength
surveys) are clearly larger than the correction made when taking into
account the additional contribution of the UV (see the dispersion in
Fig.\,2.a). This situation might change however in the near
future. More data from {\it Spitzer\,} should allow us to reduce the
uncertainties dominating our estimate of the IR energy density (e.g.,
translation from 24\mic fluxes to total IR luminosities, cosmic
variance, AGN ``contamination'').  The ``IR+UV'' view should then
provide the most accurate constraint on the star formation history.

\section*{Acknowledgments}
This work was supported by NASA through the Jet Propulsion Laboratory (subcontract
 \#960785), and 
I wish to thank my colleagues from the MIPS instrument team for their contribution to 
these results. I am also grateful to David Elbaz and Herv\'e Aussel for having
organized such a fruitful and pleasant conference.

\end{document}